\documentclass{aastex62}

\usepackage[utf8]{inputenc}
\usepackage[english]{babel}
\usepackage{amsmath}
\usepackage{amsfonts}
\usepackage{amssymb}
\usepackage{graphicx}
\usepackage{gensymb}
\usepackage{tabularx}
\usepackage{array}
\usepackage{natbib}
\setcitestyle{round}
\usepackage{hyperref}
\newcolumntype{Y}{>{\centering\arraybackslash}X}

\providecommand{\keywords}[1]
{
  \small	
  \textbf{\textit{Keywords---}} #1
}

\RequirePackage{natbib}
\renewcommand\cite{\citep}

\renewcommand{\textbf}[1]{#1}

\begin{document}

\title{\Large A massive primordial atmosphere on early Mars}

\author[0000-0001-5985-2863]{Sarah Joiret}
\affiliation{Collège de France, Université PSL, 75005 Paris, France}
\author[0000-0001-8476-7687]{Alessandro Morbidelli}
\affiliation{Collège de France, Université PSL, 75005 Paris, France}
\affiliation{Laboratoire Lagrange, Université Cote d’Azur, CNRS, Observatoire de la Côte d’Azur, Boulevard de l’Observatoire,  06304 Nice Cedex 4, France}
\author{Rafael de Sousa Ribeiro}
\affiliation{Sao Paulo State University, UNESP,Campus of Guaratingueta, Av. Dr. Ariberto Pereira da Cunha, 333 -
6 Pedregulho, Guaratingueta - SP, 12516-410, Brazil}
\author[0000-0003-0962-0049]{Guillaume Avice}
\affiliation{Université Paris Cité, Institut de physique du globe de Paris, CNRS, 75005 Paris, France}
\author[0000-0002-1462-1882]{Paolo Sossi}
\affiliation{Institute of Geochemistry and Petrology, ETH Zürich, Sonneggstrasse 5, CH-8092Zürich, Switzerland}

    \begin{abstract}
Mars finished forming while the solar nebula was still present, and acquired its primordial atmosphere from this reservoir. The absence of a detectable cometary xenon signature in the present-day Martian atmosphere suggests that the capture of solar nebular gas was significant enough to dilute later cometary contributions. By quantifying the mass of cometary material efficiently retained on Mars, we place a lower bound on the mass of the primordial Martian atmosphere. To test the robustness of our conclusions, we use cometary bombardment data from two independent studies \textbf{conducted within a solar system evolutionary model consistent with its current structure}. Our calculations show that, even under the most conservative scenario, the minimal mass of the primordial martian atmospheres would yield a surface pressure of no less than 2.9 bar. Such a massive nebular envelope is consistent with recent models in which atmospheric capture is strongly enhanced by the presence of heavier species on Mars - due to outgassing or redox buffering with a magma ocean.

    \end{abstract}
    \vspace{4mm}

\section{Introduction}
\label{Intro}
Mars formed rapidly within the first 4 million years of the solar system \citep{Dauphas2011}, while the solar nebula was still present \citep{Haisch2001, Wang2017, Weiss2021}. Heavy noble gas isotopes, which are key tracers of planetary volatile evolution \citep{Pepin1992}, indicate that Mars acquired its atmosphere from this nebular reservoir. Both in situ measurements \citep{Conrad2016} and mass spectrometry analyses of Martian meteorites \citep{Ott1988, Mathew1998, Avice2018Mars} confirm that atmospheric krypton (Kr) and xenon (Xe) have a solar origin. In contrast, the Kr isotopic composition of the Martian mantle is chondritic \citep{Peron2022Mars}, ruling out magma ocean outgassing or interior fractionation as main sources of the solar-like noble gas signature in the atmosphere. Instead, these gases were directly accreted from the solar nebula after Mars' mantle rapidly formed from chondritic material.

Sometime after the dispersal of the nebula, a dynamical instability among the giant planets scattered comets into the inner solar system \citep{Tsiganis2005, Gomes2005, Nesvorny2018review}. This cometary bombardment reshaped the volatile inventories of the terrestrial planets \citep{Joiret2023}. On Earth, cometary Xe appears in the primordial Xe signature of the atmosphere \citep{Marty2017}. Mars, however, shows no evidence of such cometary Xe signature \citep{Conrad2016, Peron2022Mars}.
This difference likely stems from Earth’s last giant impacts occurring \textit{after} solar nebula dispersal - erasing most
of the solar signatures in Earth's atmosphere - whereas Mars finished forming \textit{before} nebula dispersal \citep{Avice2018Mars}. Mars must have accreted solar nebular gas with high efficiency, as its atmospheric Xe retained a predominantly solar signature despite later cometary bombardment. Efficient capture of solar nebular gas could also account for the elevated neon (Ne) abundance inferred in the Martian mantle \citep{Kurokawa2021}. In this study, we quantify the mass of the cometary bombardment on Mars from dynamical constraints, providing a lower bound on the mass of the solar nebula that early Mars must have accreted to overwhelm any cometary Xe signature. 

\medskip

\section{Minimal mass of the primordial atmosphere on Mars}
We set the lower bound on the mass of the primordial Martian atmosphere, starting from a robust estimate of the cometary material \textit{efficiently} accreted during the bombardment episode. \textbf{This estimate is derived through a multi-step modeling framework that integrates: (1) the velocity distribution and impact probabilities of comets during the cometary bombardment phase; (2) the retention efficiency of cometary material depending on impactor sizes and velocities; (3) the size-frequency distribution and total mass of comets in the primordial trans-Neptunian disk; and (4) a correction for comet disruption effects.
The total cometary mass efficiently accreted on Mars is given by:
\begin{equation}
m_\text{comet} =  \sum_{i} M_\text{tot}(D \le 100 \text{ km}) \, p_{\text{coll}} \, f(D_i) \, \frac{\chi_i(D_i)}{\alpha \, \beta_i}
\label{eq:comet}
\end{equation} 
where each of these factors, including the total mass of comets smaller than 100 km in the primordial disk $M_\text{tot}(D \le 100 \text{ km})$, the collision probability $p_{\text{coll}}$, size distribution (fraction of the total mass in each comet size-bin $f(D_i)$), retention efficiency $\chi_i(D_i)$, obliquity correction ($\alpha$), and disruption factor ($\beta_i$) is introduced and detailed in the subsequent subsections. The last subsection presents an estimate of the minimal mass of Mars' primordial (solar-derived) atmosphere necessary to dilute cometary xenon signatures.}

\subsection{Velocity distribution and collision probability during the cometary bombardment}
\begin{figure}
	\centering
        \includegraphics[width=0.7\linewidth]{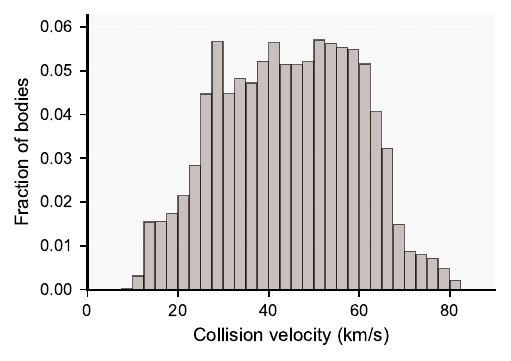}
	\includegraphics[width=0.7\linewidth]{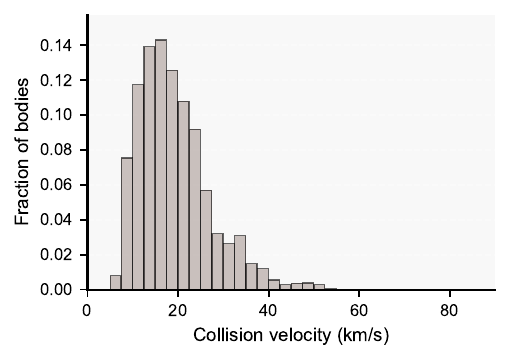}
     
	\caption{\small Distribution of collision velocities between comets and Mars in \textbf{J24 (top)} and R25 \textbf{(bottom)}. Collision velocities between each pair of bodies are weighted by the collision probability per unit of time at each possible collision orientation. \textbf{The dynamical instability is more violent in J24, which leads to higher collision velocities.}}
	\label{fig:vel_dist}
\end{figure}

We first determined the velocity distribution $f(v_j)$ (Figure \ref{fig:vel_dist}) and \textbf{average} collision probability $p_{coll}$ of comets impacting Mars during the cometary bombardment. This was achieved by combining \textit{N}-body simulations of the early evolution of the solar system with an {\"O}pik/Wetherill approach \citep{Wetherill1967, FarinellaDavis1992, Bottke1994} \textbf{- a semi-analytical method for computing collision probabilities and velocities between two bodies, based on their orbital elements and assuming random encounter geometries}. \textbf{Although the existence of a phase of dynamical instability of the giant planets, destabilizing the trans-Neptunian cometary disk, is well accepted, the dynamical character of the instability is not uniquely determined. Thus,} to assess the sensitivity of our results to its dynamical properties, we used simulation outputs from two independent studies: \citet{Joiret2024}, \textbf{featuring a violent instability,} and \citet{Ribeiro2025}, \textbf{featuring a milder instability,} hereafter J24 and R25 respectively. 

\textbf{All simulations started at the time of gas disk dispersal ($t_0$) and spanned 100 Myr. In J24, simulations were performed using GENGA \citep{Grimm2014}. Large bodies, including planetary embryos and planets, were treated with full gravity, while smaller bodies such trans-Neptunian objects were only influenced by massive objects. The primordial reservoir of trans-Neptunian objects was distributed between 21 and 30 AU, with a total mass of 25 $M_{\Earth}$. This configuration follows predictions from models of the early solar system's dynamical evolution \citep{Nesvorny2018review}, which account for collisional losses within the primordial disk \citep{Bottke2023}. The outer disk was represented by $10^4$ comets. To account for the effects of an \textit{Early Instability} scenario \citep{Clement2018}, giant planets were forced into a dynamical instability $\sim$ 2 Myr after gas disk dispersal. This was achieved using simulation outputs of giant planets evolution from \citet{Clement2021b}.}

\textbf{In R25, simulations were performed using REBOUND \citep{Rein2012}. The reservoir of trans-Neptunian objects initially extended from about 22 to 30 AU, and comprised $10^6$ massless, test particles. For computational efficiency, in the present study, we limited our analysis to 88,000 particles, corresponding to one in ten simulations from the original R25 dataset. Jupiter and Saturn were initially placed in their current orbits, while Uranus and Neptune were initially
placed inside their current orbits and then forced to migrate outward \citep{Nesvorny2017}. The initial
orbits of Uranus and Neptune were a$_U$ = 16 AU, e$_U$ = 0, i$_U$ = 0, and a$_N$ = 24 AU, e$_N$ = 0, i$_N$ = 0, respectively. The timing of the instability occurs at 10 Myr after gas disk dispersal.}

\textbf{It should be noted that the instability itself — that is, the evolution of the giant planets due to interactions with the massive outer disk — was modeled in earlier studies (e.g.  \citet{Nesvorny2012} or \citet{Clement2021b}. In J24 and R25, this interaction is not simulated directly; the resulting dynamical pathways are instead adopted from those prior models and applied to massless particles in order to study their dynamics and resulting bombardment (i.e. collision probability and velocity). Our results are not sensitive to the total mass of the primordial outer disk, but rather (as detailed below) to the size frequency distribution of comets. This is because massive objects that are not numerous enough to hit Mars with any significant probability (e.g Pluto-size bodies) don't matter in our study, even if they can carry most of the mass of the disk.}

\textbf{We considered numerical outputs from both J24 and R25 to estimate the collision rate between comets and a Mars-like planet. The algorithm used for these collision rate calculations is described in detail in \citet{Joiret2024}.}

The \textbf{average} collision probability $p_{coll}$ of comets impacting Mars during the cometary bombardment per comet initially in the trans-Neptunian disk is $\sim 10^{-6}$ in J24, and $\sim 3.1 \times 10^{-7}$ in R25. The latter is very close to the value of $3.3 \times 10^{-7}$ found by \citet{Nesvorny2023}.
The more intense cometary bombardment in J24 is related to higher collision velocities, related to a more violent giant planet instability (in a medium with a given particle density, the collision probability is proportional to the relative velocity). However, these two factors —higher impact probabilities and reduced retention efficiencies due to faster impacts- are expected to roughly compensate with each other when considering the net mass of cometary material retained by Mars. \textbf{This compensatory effect implies that our estimates are not highly sensitive to the specific dynamical scenario adopted, as shown below}. Despite different dynamical histories, both datasets yield comparable mass estimates of cometary material retained on Mars.

\subsection{Retention Efficiencies}
\begin{figure}
	\centering
	\includegraphics[width=0.7\linewidth]{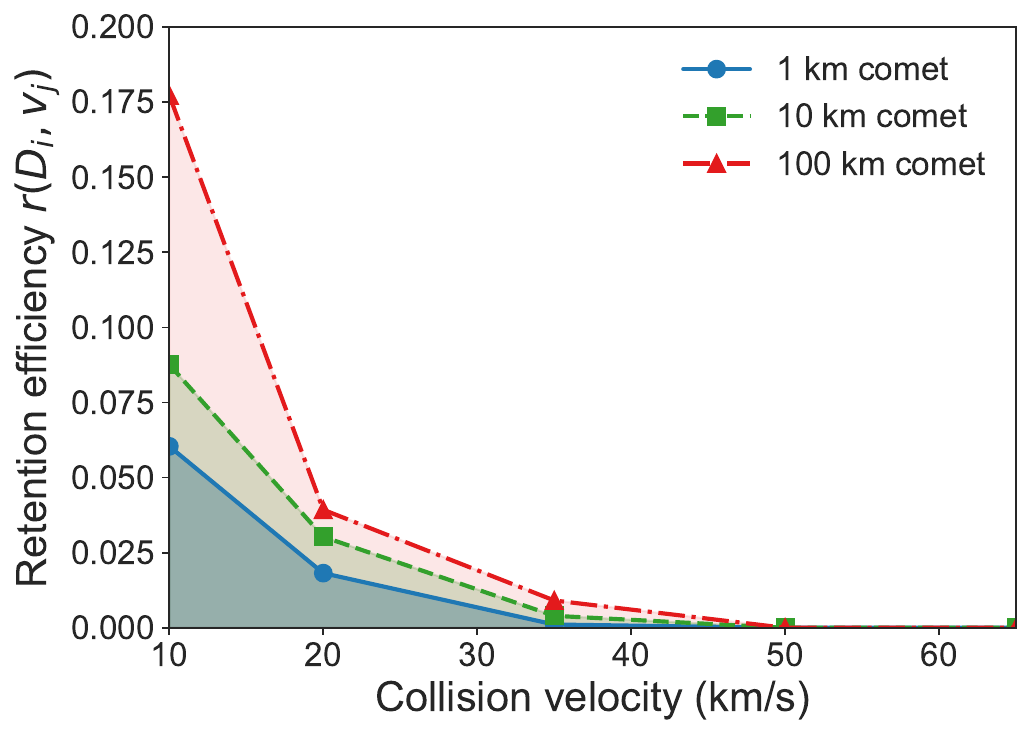}
	\caption{\small Retention efficiency of cometary material on Mars across different impact velocities and comet sizes, obtained from iSALE2D simulations.}
	\label{fig:retention}
\end{figure}

We used hydrodynamic impact simulations to calculate the retention efficiency of cometary material on Mars \textbf{(i.e. the mass of the cometary impactor that stays on the planet divided by the total mass of the impactor)}, $r(D_i,v_j)$, across a range of impactor sizes ($D_i = [1,\ 10,\ 100] \text{ km}$) and velocities ($v_j = [10,\ 20,\ 35,\ 50,\ 65] \text{ km/s}$).
Figure \ref{fig:retention} summarizes the computed retention efficiencies for different impact conditions.

\textbf{Impact simulations were performed using the iSALE2D shock physics code. The simulation domain was a two-dimensional box with cylindrical symmetry, featuring free-slip boundary conditions on the left and right, no-slip at the bottom, and an outflow condition at the top. The target was a basaltic Martian surface, modeled by the ANEOS using the default parameters in iSALE2D. Comets were assumed to be made of pure water ice, modeled by Tillotson equations of state (also using the default parameters in iSALE2D). Martian surface gravity was set at 3.72 m/s$^2$.} 

\textbf{In our 2D simulations, all impacts are vertical. However, the most probable impact angle is 45$\degree$ \citep{Pierazzo2000}; and both laboratory experiments and 3D numerical simulations show that projectile retention decreases as the impact angle (measured from the horizontal) decreases \citep{Pierazzo2003, Daly2016}. To account for this discrepancy, we introduce a correction factor $\alpha$, equal to 1.32, derived from comparative studies of cometary impacts on Mars performed both in 2D simulations and at oblique angles, including 45$\degree$ in 3D simulations \citep{Pierazzo2003, Pierazzo2006}. The final estimate of the cometary mass efficiently retained on Mars should be divided by this correction factor. Although studies from \citep{Pierazzo2003, Pierazzo2006} focus on relatively low-velocity impacts, the correction factor $\alpha$ offers a reasonable first-order estimate for angle-averaged retention efficiency. As illustrated in Figure \ref{fig:retention}, retention of cometary material on Mars is largely dominated by low-velocity impacts. Most importantly, our aim is not to precisely model angle dependence, but to provide a broad estimate of the retained cometary mass. Applying this correction does not change our results at the order-of-magnitude level, and is chosen to ensure a conservative estimate.}

\textbf{Each simulation employed a resolution of 50 cells per projectile radius, corresponding to a grid spacing of 10 m, 100 m, and 1 km for 1-km, 10-km, and 100-km impactors, respectively. 50 CPPR is a good compromise between accuracy and computational cost. Also, this is consistent with prior studies that show convergence of global metrics for resolutions of 20 CPPR or higher \citep{Pierazzo2008}.
All simulations began with the first contact between the projectile and the target and stopped after the collapse of the transient crater. We ensured that all simulations were run until the time evolution of ejected material asymptotically leveled off, indicating convergence. This plateau behavior confirms that the bulk of the impact processes had concluded and that the calculated retention efficiencies were not sensitive to longer run times. Figures that illustrate these asymptotic trends can be found in Supplementary Materials. Our initial approach was to estimate Xe loss by applying the Maxwell–Boltzmann distribution to the post-impact temperature distribution. However, peak temperatures never reached values high enough to accelerate Xe atoms beyond Mars’ escape velocity, due to the high atomic mass of Xe. So instead, retention efficiency was computed by tracking the post-impact material distribution at each timestep and measuring the fraction of cometary mass that did not reach Mars' escape velocity.} 

\textbf{We note that our retention efficiencies for small impactors (e.g., 1 km at 20 km/s, see Figure \ref{fig:retention}) are significantly lower than those reported in previous studies such as \citet{Svetsov2007}, who found that $\sim$ 90\% of the impactor mass may be retained under similar impact conditions. This discrepancy is primarily due to the fact that such studies include the presence of a dense atmosphere using the SOVA code - which offers a better treatment of ablation than iSALE. Atmospheric drag and ablation significantly enhance retention for kilometer-size impactors. In our simulations, we neglected the presence of an atmosphere, which leads to an overestimation of cometary impact velocities and underestimation of the retention efficiencies (mainly in the case of smaller impactors). This is intentional: our objective is to determine a strict lower bound on the retained mass and on the Martian primordial atmosphere. Moreover, as we show in Section \ref{sect_mass}, the cometary bombardment was dominated in mass by impactors larger than $\sim$ 10 km, for which atmospheric effects are minimal.}

\textbf{Then, we studied the cumulative retention of all accreted cometary material by calculating} the fraction $\chi_i(D_i)$ of cometary material efficiently accreted on Mars for each diameter $D_i$ of comets, given a normalized velocity distribution $f(v_j)$:
\begin{equation}
\chi_i(D_i) = \sum_{j} r(D_i,v_j) f(v_j)
\end{equation}

\begin{table}[h!]
\centering
\begin{tabular}{lccc}
\hline \hline
         & 1 km & 10 km & 100 km  \\ \hline
        \textbf{J24} & 0.31 \% & 0.58 \% & 0.89 \% \\ \hline
        \textbf{R25} & 2.41 \%  &  3.79 \% & 6.20 \%  \\ \hline
    \end{tabular}
    \caption{Fraction $\chi_i(D_i)$ of cometary material efficiently accreted on Mars for each diameter $D_i$ of comets, shown for cases J24 and R25. Case R25 demonstrates a higher retention of cometary material, due to lower collision velocities.}
    \label{table:retention}
\end{table}

\subsection{Mass of the continuous cometary bombardment}
\label{sect_mass}
\textbf{We derived the total mass of cometary material accreted on Mars from the \textit{continuous} component of the bombardment — that is, the bombardment carried by objects that are numerous enough that their collision with Mars is certain (i.e. their number times their collision probability is significantly larger than 1). It turns out that the continuous component is carried by comets smaller than 100 km in diameter. The cumulative mass they carry to Mars is expressed as} $M_\text{tot}(D \le 100 \text{ km}) \, p_{\text{coll}}$, where $M_\text{tot}(D \le 100 \text{ km})$ is the total mass of comets smaller than 100 km in the primordial trans-Neptunian disk and $p_{\text{coll}}$ is the average collision probability of comets impacting Mars following the dynamical instability. \textbf{The continuous component is complemented by the stochastic component, carried by large comets (those $>$100 km in diameter). The latter population was too rare ($\sim 10^7$ in the primordial trans-Neptunian disk, \citet{Nesvorny2017}) for their impacts to be treated as a smooth flux, given the very low collision probabilities $p_{\text{coll}} \sim 10^{-7} \text{-} 10^{-6}$. Instead, such impacts would have occurred as isolated events rather than as a continuous process.} Since our objective is to estimate a conservative lower bound on the mass contribution of the cometary bombardment to Mars' primordial atmosphere, we excluded this stochastic component from our calculations.

The total mass of comets smaller than 100 km in the primordial trans-Neptunian disk was determined using the cumulative size distribution \( N(>D) \sim 10^{12}  (\frac{D}{\text{1 km}})^{-2.1} \) from \citet{Nesvorny2017}, which was inferred from present-day populations of Jupiter Trojans (D$>$ 10 km) and Kuiper Belt Objects (D$>$ 50 km), flux of Jupiter-family comets (D$>$ 1 km) and based on the mass of the original disk needed to generate the plausible dynamical evolution of the solar system \citep{Nesvorny2012, Nesvorny2016, Deienno2017}. The corresponding differential size distribution is expressed as: 

\begin{equation}
n(D) dD \sim C \times (\frac{D}{\text{1 km}})^{-3.1} dD
\end{equation}

with $C$ calculated as \( 2.1 \times 10^{12} \). The total mass of comets with diameters below 100 km, \(M_{tot}(D\le100 \text{ km})\), is obtained by integrating over the size distribution: 

\begin{equation}
M_{tot}(D\le100 \text{ km}) = \rho \int_{0}^{100} \frac{4}{3} \pi \left( \frac{D}{2} \right)^3 n(D) dD \simeq 5 \times 10^{28} \ \text{g}
\end{equation}

assuming a cometary bulk density $\rho$ of 500 kg/m$^3$ \citep{Richardson2007, Brown2013, Groussin2019}, expressed in kg/km$^3$ in the calculations.

The mass fraction of comets within this size range in the primordial disk is then evaluated for three representative diameters ($D_i = [1,\ 10,\ 100] \text{ km}$), as follows:
\begin{equation}
f(D_i\sim 1\text{ km})=\frac{M_\text{tot}(D \le 5 \text{ km})}{M_\text{tot}(D \le 100 \text{ km})}
\end{equation}

\begin{equation}
f(D_i\sim10\text{ km})=\frac{M_\text{tot}(5 < D \le 50 \text{ km})}{M_\text{tot}(D \le 100 \text{ km})} 
\end{equation}

\begin{equation}
f(D_i\sim100\text{ km})=\frac{M_\text{tot}(50 < D \le 100 \text{ km})}{M_\text{tot}(D \le 100 \text{ km})}
\end{equation}

Results are summarized in Table \ref{table:masscom}.

\begin{table}[ht!]
\centering
\begin{tabular}{lccc}
\hline \hline
        \rule{0pt}{4ex} \textbf{M$_{tot}$($D\le$100 km)} & \textbf{f($D_i\sim$1km)} & \textbf{f($D_i\sim$10km)} & \textbf{f($D_i\sim$100km)} \\
        & \rule{0pt}{4ex}$=\frac{M_\text{tot}(D \le 5 \text{ km})}{M_\text{tot}(D \le 100 \text{ km})}$ \vspace{1ex} & \rule{0pt}{4ex}$=\frac{M_\text{tot}(5 < D \le 50 \text{ km})}{M_\text{tot}(D \le 100 \text{ km})}$ \vspace{1ex} & \rule{0pt}{4ex}$=\frac{M_\text{tot}(50 < D \le 100 \text{ km})}{M_\text{tot}(D \le 100 \text{ km})}$ \vspace{1ex} \\ \hline
         \rule{0pt}{1ex} $\sim 5 \times 10^{28}$ g & 6.6 \% & 46.7 \% & 46.7 \% \\ \hline
    \end{tabular}
    \caption{Total mass $M_\text{tot}(D \le 100 \text{ km})$ of comets smaller than 100 km and mass fraction $f(D_i)$ of comets of diameter $D_i$ in the primordial trans-Neptunian disk (considering objects smaller than 100 km only), as calculated from the cumulative size distribution of primordial trans-Neptunian objects in \citet{Nesvorny2017}.}
    \label{table:masscom}
\end{table}


\subsection{Comet disruption}
Comets tend to lose mass due to long-lasting activity and spontaneous disruption \citep{Chen1994, Levison2002}. Alternatively, they can become dormant due to the formation of a refractory crust that inhibits dust and volatile emissions \citep{Coradini1997, Ye2016}. If spontaneous comet disruption is taken into account, \citet{Nesvorny2023} showed that the impact flux of comets on Mars is reduced by a factor $\beta_\text{i}$ =  36, 4.9 and 1.7 for comets of 1, 10 and 100 km in diameters, respectively.

\subsection{Minimal mass of the primordial atmosphere on Mars}
\textbf{Finally, by combining all the components described above, we obtain the expression for the total cometary mass efficiently accreted on Mars  (Equation~\eqref{eq:comet}):}
\[
m_\text{comet} =  \sum_{i} M_\text{tot}(D \le 100 \text{ km}) \, p_{\text{coll}} \, f(D_i) \, \frac{\chi_i(D_i)}{\alpha \, \beta_i}
\]

Taking into account comet disruption, we found $m_\text{comet} = 1.1 \times 10^{20}$ g in the J24 case, and $m_\text{comet} = 2.7 \times 10^{20}$ g in the R25 case. Neglecting it (i.e. setting $\beta_i = 1$), we found $m_\text{comet} = 2.6 \times 10^{20}$ g in the J24 case, and $m_\text{comet} = 5.4 \times 10^{20}$ g in the R25 case.

\begin{figure}
	\centering
	\includegraphics[width=0.7\linewidth]{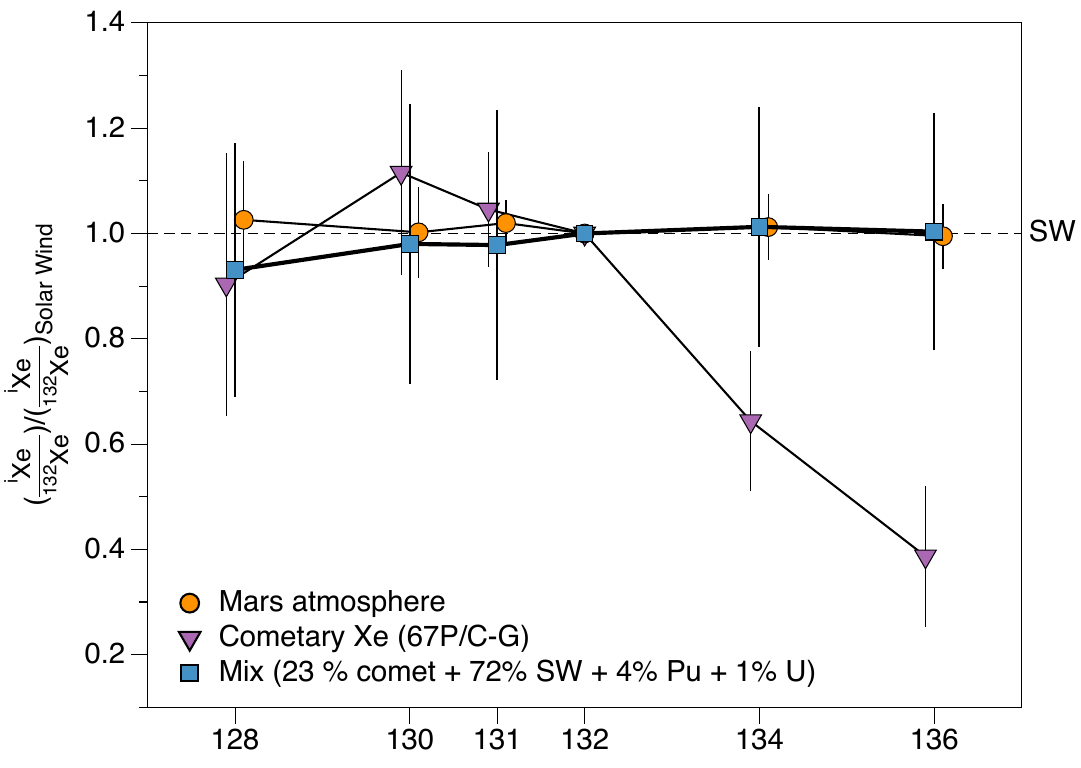}
	\caption{\small A mixture of solar and cometary Xe as the origin of the primordial atmospheric Xe on Mars. The Xe isotopes are normalized to $^\text{132}\text{Xe}$ and the solar wind composition SW-Xe (the horizontal dashed line). SW-Xe data are from \citet{Meshik2020}, Cometary (67P/Churyumov-Gerasimenko) data are from \citet{Marty2017} and Martian atmosphere data are from \citet{Conrad2016}, corrected for mass-dependent fractionation. Xe production due to Pu-fission and U-fission is taken into account. Error bars are 1$\sigma$.}
	\label{fig:Mars_com}
\end{figure}

It remains possible that the Xe isotopic signature of the Martian atmosphere comprises a minor cometary contribution \citep{Swindle1997, Marty2016}. Using a Linear Least Squares fit combined with Monte Carlo error propagation ($10^6$ random iterations), we calculated that the Martian atmospheric xenon signature may comprise up to 16\% cometary Xe for 82\% solar Xe, accounting for the substantial measurement uncertainties in 67P/Churyumov-Gerasimenko. Forcing elevated rates of Xe production due to Pu-fission (e.g., $\sim$ 4\%) increases the inferred cometary contribution (23\% cometary Xe for 72\% solar Xe, Figure \ref{fig:Mars_com}). Although there is no apparent evidence for Xe production due to Pu-fission \citep{Conrad2016, Avice2018Mars}, this endmember scenario serves as an absolute lower bound on the solar-derived atmospheric component necessary to erase a cometary signature. Under this constraint, the solar-to-cometary Xe ratio is $\frac{72}{23} \simeq 3.1$.

The minimum mass of Mars' primordial atmosphere is thus the amount of solar nebular gas required to get at least 3.1$\times$ more solar Xe than cometary Xe. Given the fact that there is $\sim$ 100$\times$ more mass in the gaseous component of the solar nebula (mainly hydrogen and helium) than in its condensable (or dust) component \citep{Pascucci2010}, and that the Xe to condensable species ratio is 3$\times$ lower in comets than in the solar nebula \citep{Marschall2025}, we have: 

\begin{align}
m_\text{atm, min} &= m_\text{gas, solar} = 100 \times m_\text{cond, solar} = 100 \times m_\text{Xe, solar} \times \left( \frac{m_\text{cond}}{m_\text{Xe}} \right)_\text{solar} \nonumber \\ 
&\geq 100 \times 3.1 \times m_\text{Xe, comet} \times \frac{1}{3} \times \left( \frac{m_\text{cond}}{m_\text{Xe}} \right)_\text{comet} = 103.3 \: m_\text{cond, comet}
\end{align}

where $m_\text{atm, min}$ is the minimum mass of the primordial Martian atmosphere, $m_\text{gas, solar}$ is the mass of solar nebula gas that was accreted on Mars and $m_\text{cond, solar}$ is the corresponding mass of condensable species in the solar nebula \textbf{(that is, all species other than hydrogen and helium)}, $m_\text{cond, comet} \sim m_\text{comet}$ denotes the cometary mass, and $m_\text{Xe, solar}$ and $m_\text{Xe, comet}$ are the masses of Xe in the solar nebula and in comets, respectively.

Using the derived values, and taking into account comet disruption, the \textbf{minimum mass of the primordial Martian atmosphere} is estimated to be between approximately \( 1.1 \times 10^{22} \) g and \( 2.8 \times 10^{22} \) g, in the J24 and R25 cases, respectively, corresponding to surface pressures of $2.9$ bar and $7.3$ bar on Mars. Neglecting comet disruption, the minimal mass is estimated to be between approximately \( 2.7 \times 10^{22} \) g and \(5.6 \times 10^{22} \) g, in the J24 and R25 cases, respectively, corresponding to surface pressures of $6.9$ bar and $14.5$ bar.

\section{Discussion}
To reconcile the estimated Ne abundances in the Martian mantle, \citet{Kurokawa2021} determined that the early Martian atmosphere would have required a Ne partial pressure of $\sim$ 10 Pa. Assuming a $^{20}$Ne molar fraction of 1.8 $\times$ $10^{-4}$ in the solar nebula, this corresponds to a total atmospheric pressure of $\sim$ $0.5$ bar (5 $\times$ $10^4$  Pa) and an atmospheric mass of 2 $\times$ $10^{21}$ g. This estimate aligns with hydrodynamic simulations of solar nebula capture, which predict a primordial Martian atmosphere of at most 1 bar \citep{Stokl2015}. Our lower bound extends beyond this upper limit. Recent work showed that mixing an outgassed component (e.g., CO$_2$-rich) to the primordial H$_2$-rich atmosphere enhances the nebular captured gas inventory by $\sim$ 1–3 orders of magnitude \citep{Pahlevan2025}. In particular, for planetesimal accretion rates of 0.01–1 Mars masses/Myr (corresponding to different planetary luminosities) and heavy gas inventories equivalent to 10–1000 bars of CO$_2$ at the planetary surface \citep{Elkins2008}, the captured nebula inventory is equivalent to $\sim$ 3–300 bars of H$_2$ at the surface. Such a hybrid primordial atmosphere is emerging as a compelling scenario, as it reproduces the noble gas isotopic composition observed in the Martian atmosphere \citep{Pahlevan2025}, including that of xenon.

An alternative, and arguably more natural, mechanism for increasing nebular gas retention involves H$_2$O vapor in equilibrium with a magma ocean. At the oxygen fugacity of the Martian mantle (\textbf{IW $= 0 \pm 1$}, \citet{Wadhwa2008}), redox buffering between the atmosphere and interior imposes a \textbf{\(\log(f_{\text{H}_2\text{O}} / f_{\text{H}_2})\) ratio of 0 $\pm$ 0.5 at 2000 K}, raising the mean molecular weight \textbf{from 2 to 10 $\pm$ 4 g/mol}. This increase alone enhances the captured mass of nebular gas by up to an order of magnitude—potentially yielding 10–20 bars of H$_2$ at the surface. These values are in excellent agreement with those inferred from our method. \textbf{This scenario invokes a (partial) magma ocean, which implies some exchange between the Martian atmosphere and interior. To preserve a chondritic Kr and Xe signature in the mantle, the amount of these gases dissolved during exchange must have been small - likely due to their low solubility relative to lighter noble gases \citep{Carroll1994}. In contrast, the higher solubility of He and Ne (despite their lower polarizabilities) makes their partial equilibration with the atmosphere more plausible \citep{Kurokawa2021}.}

If the entire primordial atmosphere was lost through hydrodynamic escape within a few millions years \citep{Erkaev2014}, this would impose stringent constraints on the timing of cometary bombardment, as the isotopic signatures delivered by comets could no longer be diluted once the solar primordial atmosphere is gone. \textbf{However, because atmospheric escape preferentially removes light species such as hydrogen and helium, heavier noble gases like xenon may persist longer. In this case, a residual solar Xe component could still buffer the isotopic signature of cometary Xe delivered after most of the atmosphere had been lost.}

Alternatively, if the primordial atmosphere partially persisted over prolonged timescales, our results carry important implications for early Martian climate. The persistence of H$_2$-rich gases would represent a potent greenhouse contribution, potentially sustaining warmer surface conditions during the pre-Noachian era \citep{Ramirez2014, Wordsworth2017, Hayworth2020, Turbet2021}.

\textbf{It should be noted that comets have much higher noble gas–to–H$_2$O, –C, and –N ratios than asteroids \citep{Balsiger2015, Bekaert2020}, allowing them to deliver substantial noble gases while contributing little H, C, or N compared to asteroids. This pattern is also seen on Earth, where the cometary contribution is negligible for water but significant for Ar, Kr, and Xe \citep{Marty2016}. While isotopic signatures of He, Ne, Ar, and Kr are less diagnostic than xenon, existing data suggest a cometary contribution to Mars’s atmosphere remains plausible also for these elements. \citep{Owen1995, Marty2016}.}

\section{Conclusions}
By combining noble gas data with N-body hydrodynamic impact simulations, we have quantified the minimal mass of Mars' primordial atmosphere necessary to retain a predominantly solar noble gas signature despite subsequent cometary impacts.
Our findings imply that Mars accreted a substantial gaseous envelope from the solar nebula, with minimum surface pressures ranging from 2.9 to 14.5 bar — depending on the intensity of cometary bombardment and whether comet disruption is accounted for. \textbf{These estimates are consistent with previous studies showing that the presence of heavier molecular species can significantly enhance nebular gas retention by increasing the mean molecular weight of the atmosphere \citep{Kurokawa2021, Pahlevan2025}.}
Whether Mars partially retained or rapidly lost its primordial atmosphere yields distinct interpretations. \textbf{A rapid loss would limit the dilution from late-arriving comets, though a residual solar Xe component -less affected by atmospheric loss than hydrogen- may have persisted.} In contrast, sustained retention of H$_2$-rich gases would have provided a substantial greenhouse warming, with profound implications for the pre-Noachian climate.


\section{Acknowledgements}
We gratefully acknowledge David Nesvorný and Martin Turbet for their valuable insights, which greatly contributed to the advancement of this work. Computer time for this study was partly provided by the computing facilities MCIA (Mésocentre de Calcul Intensif Aquitain) of the Université de Bordeaux and of the Université de Pau et des Pays de l'Adour, France. Numerical computations were also partly performed on the S-CAPAD/DANTE platform, IPGP, France. This project has received funding from the European Research Council (ERC) under the European Union’s Horizon Europe research and innovation program (grant agreement no. 101041122 to G.A.). RR thanks the scholarship granted from the Brazilian Federal Agency for Support and Evaluation of Graduate Education (CAPES), in the scope of the Program CAPES-PrInt (Proc~88887.310463/2018-00, Mobility number 88887.572647/2020-00, 88887.468205/2019-00). This research was supported in part by the São Paulo Research Foundation (FAPESP) through the computational resources provided by the Center for Scientific Computing (NCC/GridUNESP) of the São Paulo State University (UNESP). RR also acknowledges support provided by grants FAPESP (Proc~2016/24561-0) and by Sao Paulo State University (PROPe~13/2022).

\medskip
\bibliography{reference}

\end{document}